\documentclass{article}

\usepackage[preprint]{neurips_2025} 

\usepackage[utf8]{inputenc} 
\usepackage[T1]{fontenc}    
\usepackage{hyperref}
\hypersetup{colorlinks,allcolors=black}
\usepackage{url}            
\usepackage{booktabs}       
\usepackage{amsfonts}       
\usepackage{nicefrac}       
\usepackage{microtype}      
\usepackage{xcolor}         

\usepackage{epsfig}
\usepackage{subfigure}
\usepackage{lettrine}
\usepackage{mathrsfs}
\usepackage{subfigure}
\usepackage{graphicx}
\usepackage{diagbox}

\usepackage{booktabs} 
\usepackage{multirow}
\usepackage{stmaryrd}
\usepackage{mathrsfs}
\usepackage{pifont}
\usepackage{babel}
\usepackage{amsmath}
\usepackage{amsthm} 
\newtheorem{theorem}{Theorem}
\newtheorem{definition}{Definition}
\usepackage{natbib}

\title{RH: An Architecture for Redesigning Quantum Circuits on Quantum Hardware Devices}

\author{
  Runhong He$^{1*}$, Ji Guan$^{1}$, Xin Hong$^{1}$,  Guolong Cui$^{2}$, \AND Shengbin Wang$^{3}$, Shenggang Ying$^{1**}$ \\
  \\
  $*$ \texttt{runhong93@qq.com}\\
  $**$ \texttt{sgying@ios.ac.cn}\\ 
  \\
  1. Key Laboratory of System Software (Chinese Academy of Sciences) and State Key Laboratory \\ of Computer Science 
  Institute of Software, Chinese Academy of Sciences, 
  Beijing 100190, China  \\
  2. Arclight Quantum Co., LTD. Chinese Academy of Sciences, Beijing	101408, China\\
  3. China Telecom Quantum Information Technology Group Co., LTD, HeFei 233000, China}

\begin{document}

\maketitle

\begin{abstract}

In this paper we present an architecture that enables the redesign of large-scale quantum circuits on quantum hardware based on the entangling quantum generative adversarial network (EQ-GAN). Specifically, by prepending a random quantum circuit module to the standard EQ-GAN framework, we extend its capability from quantum state learning to unitary transformation learning. The completeness of this architecture is theoretically proved. Moreover, an efficient local random circuit is proposed, which significantly enhances the practicality of our architecture. For concreteness, we apply this architecture to three crucial applications in circuit optimization, including the equivalence checking of (non-) parameterized circuits, as well as the variational reconstruction of quantum circuits. The feasibility of our approach is demonstrated by excellent results in both classical and noisy intermediate-scale quantum (NISQ) hardware implementations. We believe our work will facilitate the implementation and validation of the advantages of quantum algorithms. 

\end{abstract}

\section{Introduction}
Quantum algorithms can provide super-polynomial or even exponential
speedups compared to their classical counterparts in solving certain
important problems \cite{qc_nielsen,qc_bb}, such as the large integers factoring \cite{shor}, unstructured database searching \cite{grover_0,grover_3} and linear systems solving \cite{hhl,wangshengbin_hhl_poisson_equations}. 
Quantum algorithms are commonly represented as a certain quantum circuit comprised of a series of quantum gates. In the initial stage of quantum algorithm design, researchers tend to focus on functional specification, without giving significant consideration to actual implementation. Hence the resultant draft circuits may contain some logic operations, whose unitaries are known but whose implementations are not explicitly defined, such as the Oracles raised in the  Grover's algorithm \cite{grover_0,grover_3}. Therefore, these high-level logic operations must be translated into the device's native gate library for respecting hardware constraints -- a process known as quantum logic circuit synthesis  \cite{circuit_synthesis_review}. In addition to manually performing it, which is a rather labor intensive and skill-demanding task in practice, some established techniques can also automatically offer workable results, such as unitary decomposition \cite{SK_algorithm,cs_decompose}, genetic algorithm \cite{genetic_algorithm_for_logic_circuit_synthesis}, evolutionary algorithm \cite{evolutionary_algorithm_for_oracle}, deep reinforcement learning \cite{gate_decompose_prl_dqn, he_DRL}, and variational quantum algorithm \cite{gate_decompose_with_VQA,gate_decompose_with_VQA_adaptive_circuit_compression, he_MS}.
 
%

Depending on their performance, various approaches might yield different results, resulting in the need to check the equivalence between their outputs. There are numerous algorithms that can do this, such as the decision graph \cite{Decision_diagram_for_circuit_equivalence} for non-parameterized circuits and the ZX calculus \cite{ZX_for_circuit_equivalence} for parameterized circuits. Unfortunately, their chief innovations might fail to work in certain cases, leaving the classical simulation to become the last resort. Suffering from the inefficient classical simulation of quantum systems this strategy scales very poorly with respect to the system's size, and turn impractical as the number of qubits gets large. Thus it is highly desirable to establish an algorithm that do not suffer performance degradation as the involved qubits increase.

These problems are classically hard but quantumly easy. Reference~\cite{QGAN_2} proposes the entangling quantum generative adversarial algorithm (EQ-GAN) to reproduce a given reference quantum state based on quantum hardware \cite{QGAN_1}. It circumvents the classical simulation of quantum systems, and thereby can be applied to a large scale. The EQ-GAN can always converge to a provably optimal Nash equilibrium and avoids the mode collapse which may occur in traditional Q-GAN proposals \cite{QGAN_in_superconding_system,QGAN_0}. In addition, the EQ-GAN permits mitigating uncharacterized coherent error due to miscalibrated gate parameters. However, it cannot be directly used to redesign quantum circuits, because it actually learns only the first column of the corresponding unitary matrix. In our requirements, it is necessary to learn all elements of the unitary matrix.
\begin{definition}
	We define quantum circuit redesign as the task of transforming a given quantum circuit into an alternative circuit architecture while preserving its fundamental functionality.
\end{definition}
Many important applications fall naturally into the domain of quantum circuit redesign, such as the equivalence checking of quantum synthesized circuits, quantum circuits optimization and gate decomposition.

In this paper, we propose a novel approach which permits researchers to redesign large-scale quantum circuits on quantum hardware. We refer to this approach as the RH (Redesign circuits on quantum Hardware) architecture for simplicity. Our RH architecture contains and extends the EQ-GAN \cite{QGAN_2,QGAN_1}. Specifically, by introducing an additional random quantum circuit module, the RH architecture can be directly applied to checking the equivalence between different circuits, whether they are parameterized or non-parameterized. Furthermore, this approach remains valid for variational learning of reference circuit functionality by structurally distinct parameterized circuits, owing to its inherent compatibility with quantum machine learning frameworks \cite{QML_review_QST}. The completeness of this architecture is theoretically proven. Moreover, an efficient local random circuit has been developed, which significantly enhances the practicality of this framework. To validate the RH architecture, we exemplarily implement three applications both in classical simulation and quantum hardware. Collectively these applications highlight the excellent performance of the RH architecture.



Our work opens a venue for redesigning quantum circuits using quantum hardware, permitting a wide range of applications such as large-scale quantum circuit optimization, equivalence checking, etc. It is important to highlight that our RH architecture can also be incorporated into other techniques for further advanced applications. For instance, it can be used to replace the classical simulation as the last resort, which is often the computational bottleneck, in some algorithms such as ZX \cite{ZX_for_circuit_equivalence} and decision diagram \cite{Decision_diagram_for_circuit_equivalence} for circuit simplification or optimization. We believe our RH architecture will facilitate the implementation and validation of advanced quantum algorithms.
\section{Preliminaries}
For readers' convenience, this section introduces some quantum computation concepts central to our work. For additional quantum computing fundamentals, we refer readers to Reference~\cite{qc_nielsen}.

\subsection{Quantum State and Quantum Circuit}
In quantum computation, the fundamental unit of information is the quantum bit (qubit) - the quantum counterpart of the classical bit.  The qubit's basis states $|0\rangle= [1, 0]^T$ and $|1\rangle=[0, 1]^T$ correspond to the classical $0$ and $1$ states, respectively. The essential quantum mechanical distinction is that a qubit can exist in any superposition of these basis states, described by: $|\psi\rangle=\alpha|0\rangle + \beta|1\rangle=[\alpha,\beta]^T$, where $\alpha$ and $\beta$ are complex probability amplitudes obeying the normalization condition $|\alpha|^2+|\beta|^2=1$. When measured, the qubit's superposition state collapses randomly to either $|0\rangle$ or $|1\rangle$, yielding  either outcome $0$ (with probability $|\alpha|^2$) or outcome $1$ (with probability $|\beta|^2$). In Dirac notation, the bra state $\langle\psi|$ is defined as the Hermitian conjugate of the ket state $|\psi\rangle$, i.e., 
$\langle\psi|=(|\psi\rangle)^\dagger=\begin{pmatrix}\alpha^*,\beta^*\end{pmatrix}$.
In an $n$-qubit system, the quantum state is spanned by a set of $2^n$ basis states. These basis states can be expressed as:
$|x_1x_2\cdots x_n\rangle\equiv|x_1\rangle\otimes|x_2\rangle\otimes\cdots\otimes|x_n\rangle,$
where $x_i\in\{0,1\}$, and $\otimes$ denotes the tensor product.

The overlap between two quantum states can be quantified by the fidelity $F=|\langle\psi'|\psi\rangle|^2$, which ranges from 0 (orthogonal states) to $1$ (equivalent states). If two quantum states differ only by a global phase factor, namely,  $|\phi\rangle=\text{e}^{i\theta}|\psi\rangle$ (where $\theta$ is a real phase angle), they are physically indistinguishable in all observable measurements and thus can be considered equivalent. 

Quantum states are manipulated by quantum gates represented by unitary matrices $U$ (where $U^{\dagger}U=I$). Any quantum logic operation can be realized by a circuit composed of arbitrary single-qubit rotation gates and at least one entangling two-qubit gate (e.g. $CX$ or $CZ$).

\subsection{SWAP Test}
\begin{figure}[h] 
	\subfigure[]{\includegraphics[width=0.38\linewidth]{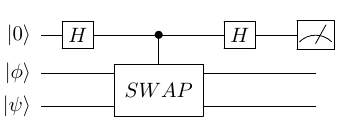}}
	\subfigure[]{\includegraphics[width=0.62\linewidth]{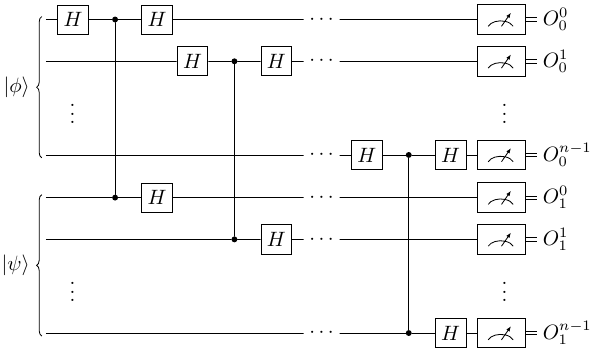}}
	\caption{\label{fig_swap_test} (a) The SWAP test circuit. (b) The destructive SWAP test circuit.}
\end{figure}
In the SWAP test \cite{swap_test}, whose circuit is visualized in Figure~\ref{fig_swap_test} (a), the measurement probability $p_1$ of obtaining outcome $1$ on the ancillary qubit reflects the difference between input states $|\phi\rangle$ and $|\psi\rangle$ through the relation:
\begin{equation}\label{p_1}
	p_1=\frac{1}{2}-\frac{1}{2}|\langle\phi|\psi\rangle|^2=\frac{1}{2}(1-F).
\end{equation}
We define the probability of test failure as 
\begin{equation}
	\label{p_failure}
	p_{\text{failure}}=2p_1=1-F.
\end{equation}
If the input states are equivalent, $p_{\text{failure}}=0$; whereas for orthogonal states, $p_{\text{failure}}$ reaches its maximum value of $1$.

By simplifying the quantum circuit of the SWAP test, we can derive its destructive variant \cite{destructive_swap_test} (Figure~\ref{fig_swap_test} (b)). Here, the term ``destructive'' emphasizes the measurement-induced collapse of all data qubits' superposition states. The destructive SWAP test incurs no practical disadvantage compared to the standard version, as both protocols inherently consume the input states during measurement.
In the destructive SWAP test, the parameter $p_1$ from the standard SWAP test is redefined as the probability of obtaining even results of $\sum_{i}O_{0}^{i}\& O_{1}^{i}$ on measurement outcomes, with symbol $\&$ referring
to the AND operation \cite{destructive_swap_test}.
\subsection{EQ-GAN}
\begin{figure*}[]
	\subfigure[]{\includegraphics[width=0.43\linewidth]{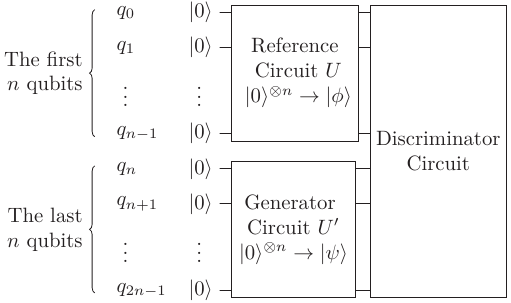}}
	\subfigure[]{\includegraphics[width=0.57\linewidth]{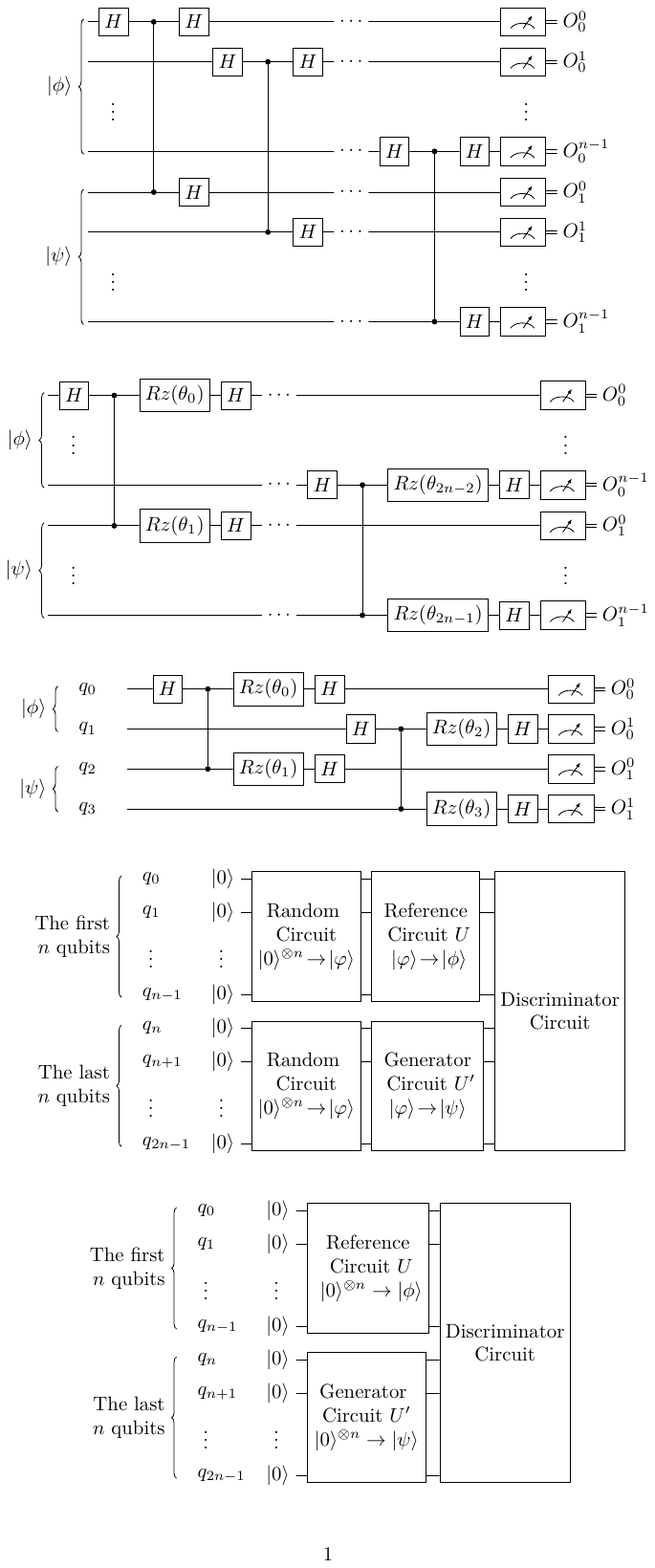}}
	\caption{\label{fig_eq_gan} (a) The overall framework of the EQ-GAN algorithm \cite{QGAN_2}. (b) The construction of discriminator in EQ-GAN \cite{QGAN_2,QGAN_1}. }
\end{figure*}
The EQ-GAN algorithm \cite{QGAN_2, QGAN_1} (Figure~\ref{fig_eq_gan} (a)) is designed to reproduce a target quantum state $|\phi\rangle$ prepared by a reference circuit $U$ on a parameterized generator circuit $U'$ through variationally learning. A discriminator circuit evaluates the overlap between the target state $|\phi\rangle$ and the generator circuit's output state $|\psi\rangle$.

The destructive SWAP test can provide a perfect metric for
states comparison in noise-free environments. Whereas in actual experiment, the two-qubit $CZ$ gate on superconducting platform suffers from unstable coherent error due to miscalibrated gate parameters, which
oscillate over the timescale of $O(10)$ minutes \cite{google_2019}.
To address this issue, EQ-GAN employs a parameterized destructive SWAP test as its discriminator, with the corresponding circuit implementation shown in Figure~\ref{fig_eq_gan} (b).
The coherent errors in $CZ$ gates can be effectively compensated by subsequent single-qubit $R_{z}(\theta_i)$ gates with variationally optimized rotation angles $\theta_{i}$ \cite{rz_for_cz_gate}, achieving high-fidelity SWAP test implementation in noisy environments.
The generator circuit is then trained to minimize the discriminator's test failure probability until $p_{\text{failure}}\rightarrow 0$, indicating that the generator's output state $|\psi\rangle$ becomes equivalent to the target state $|\phi\rangle$ prepared by the reference circuit.

\section{Proposed Method}

\subsection{Framework and Workflow \label{workflow}}

\begin{figure*}[h] 
	\centering
	\includegraphics[width=0.65\linewidth]{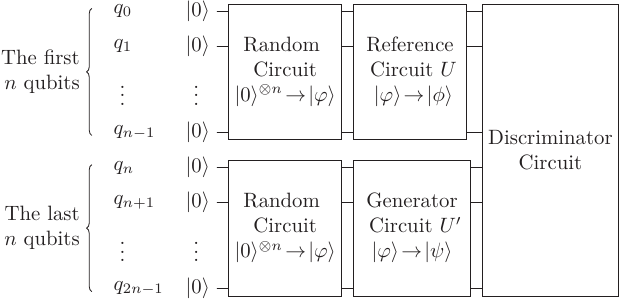}
	\caption{\label{overall_circ} The schematic diagram of the RH architecture. }
\end{figure*}

Our RH architecture is derived from the EQ-GAN \cite{QGAN_2} and
primarily consists of five components, as shown in Figure~\ref{overall_circ}:
two identical random circuits, one reference circuit $U$, one generator
circuit $U'$ and one discriminator circuit.

The heart of quantum circuit redesign lies in reproducing the functionality
of a given $n$-qubit reference circuit by a $n$-qubit generator
circuit with a new structure. To accomplish this, RH architecture need a $2n$-qubit quantum system, where the reference circuit is deployed in the first
$n$ qubits while the generator circuit is located in the last $n$ qubits. They transform the input random states $|\varphi\rangle$ prepared by two
identical random circuits into the states $|\phi\rangle$ and $|\psi\rangle$
respectively, and then feed them to the discriminator circuit. The
discriminator circuit \cite{QGAN_2} used here is the parameterized  destructive SWAP test \cite{destructive_swap_test} (Figure~\ref{fig_eq_gan} (b)),
which quantifies the overlap between its input states $|\phi\rangle$
and $|\psi\rangle$. Specifically, the probability of test failure
$p_{\text{failure}}$ equals to the infidelity $1-\left|\langle\phi|\psi\rangle\right|^{2}$, according to Equation~(\ref{p_failure}). Thus, $|\phi\rangle$ is equivalent to $|\psi\rangle$  if $p_{\text{failure}}=0$. In this case, we say the states pass the discriminator's test; otherwise, they fail the verification.

Our RH architecture begins by training the discriminator's parameters
$\boldsymbol{\theta}$ with two identical input states, i.e., $|\phi\rangle=|\psi\rangle$,
aiming at minimizing the cost function $p_{\text{failure}}$. When
$p_{\text{failure}}=0$ the discriminator could act as a perfect destructive
SWAP test under the influence of coherent error. Considering
that the unique global optimum of the discriminator is to utterly
eliminate the coherent error \cite{QGAN_2} and any input states will
lead to the same result, the zero state $|0\rangle^{\otimes n}$ is
suitable for employment as the input states in training to reduce
overhead. In other words, the random circuits, the reference circuit
and the generator circuit can be ignored in this step.

Once a discriminator with well performance has been obtained after
training, we can employ it to evaluate the overlap between the states generated by the reference and the generator circuits respectively. If the generator circuit is equivalent to the reference circuit, for any identical input $|\varphi\rangle$, their outputs should also be equivalent, and then pass the discriminator's test with $p_{\text{failure}}=0$, i.e., 
\begin{equation}
	p_{\text{failure}}\!=\!1\!-\!\left|\langle\phi|\psi\rangle\right|^{2}\!=\!1\!-\!\left|\langle\varphi|U^{\dagger}U'|\varphi\rangle\right|^{2}\!=\!1\!-\!\left|\langle\varphi|\varphi\rangle\right|^{2}\!=\!0.
\end{equation}
Otherwise, $p_{\text{failure}}$ will also reflect the overlap between two circuits, according to Equation~(\ref{p_failure}).

\subsection{Completeness and Limitations}

We define the difference matrix to capture the error
of the new circuit $U'$ with respect to the target circuit $U$,
which takes the form $D=U^{\dagger}U'$. The matrix $D$ is also an
unitary matrix, and will be an identity matrix $\mathbb{I}$ if $U=U'$, i.e., these two circuits are equivalent. A non-identity $D$ results from erroneous compilation of quantum gates in the new circuit $U'$. 

In this paper, we employ random quantum states $|\varphi\rangle$ to
conduct tests. According to the following Theorem \ref{theorem1}, for several random states $|\varphi\rangle$, if the tests always
yield the conclusion of equivalence between two circuits, i.e., 
\begin{equation}\label{eq_D_f}
	1=\left|\langle\varphi|U^{\dagger}U'|\varphi\rangle\right|^{2}=\left|\langle\varphi|D|\varphi\rangle\right|^{2}=\left|\langle\varphi|\varphi\rangle\right|^{2},
\end{equation}
the probability of that they are non-equivalence
is statistically zero. 

\begin{theorem} \label{theorem1}
	If $D|\varphi\rangle = |\varphi\rangle$ is satisfied for arbitrary random state $|\varphi\rangle$, $D$ must be the identity matrix $\mathbb{I}$.
\end{theorem}
\begin{proof}
	Since $D|\varphi\rangle = |\varphi\rangle$, it follows that 
	\begin{equation}
		\sum_{j}D_{i,j}|\varphi\rangle_{j}=|\varphi\rangle_{i},
	\end{equation}
	or 
	\begin{equation} \label{equation}
		(D_{i,i}-1)|\varphi\rangle_i=-\sum_{j\neq i}D_{i,j}|\varphi\rangle_j.
	\end{equation}
	Given that $|\varphi\rangle$ is a random state with mutually independent components (disregarding the global factor), both sides of Equation~(\ref{equation}) evaluate to $0$. 
	Therefore, $D_{i,i}=1$ and $D_{i,j}=0$ for any $j\neq i$, which implies that $D$ is an identity matrix $\mathbb{I}$.
\end{proof}

Clearly, any subtle distortion in the different matrix  relative to the identity matrix will reduce the fidelity between the output states of the reference and generator circuits, consequently yielding a non-zero $p_{\text{failure}}$ through Equations~(\ref{p_failure})  and (\ref{eq_D_f}).
Therefore, the strategy of using random quantum states as inputs for both the reference and generator circuits, with equivalence checked by the discriminator, constitutes a theoretically complete verification framework.
Another important reason for the adoption of random quantum states is that
they lead to an open-loop training style for the subsequent generator optimization, reducing the
susceptibility of the optimization to local optima akin to the usage
of random samples in the training landscape of classical neural networks
\cite{he_MS, deep_learning_book}.

It should also be noted that as the number of qubits increases, the averaged probability amplitude of random quantum states decays exponentially across computational basis states. Consequently, in certain unfavorable scenarios the error detection probability becomes exponentially small. 
Assuming $D$ includes only one non-trivial operation $U_{s}$,
one of the worst-case scenarios occurs when this operation acts on the first qubit while being controlled by the remaining
$n-1$ qubits. The resulting difference matrix takes the form $D=\mathbb{I}_{2}^{\oplus (n-1)} \oplus U_{s}$, where $\oplus$ denotes the direct sum operation of matrices.
In this case, the probability that the error goes undetected within $m$ measurement shots is about $e^{-m/2^{n-1}}$.
However, for most cases, RH can detect the existence of errors within just a few thousand shots, as we will demonstrate in subsequent experiments, making it a valuable approach compared to classical simulation methods that fail to produce reference results for large-scale circuits within reasonable time.
Furthermore, the RH framework can only probabilistically yield false negative conclusions (failing to detect actual differences between circuits) but is fundamentally impossible to produce false positives (where equivalent circuits are erroneously judged as non-equivalent) -- assuming ideal operations and perfect measurements in experiment.

\subsection{Efficient Local Random Circuit}
\begin{figure}[h] 
	\centering
	\includegraphics[width=0.35\linewidth]{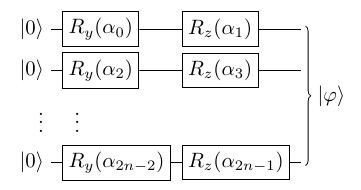}
	\caption{\label{fig:11} The local random circuit used in this paper.}
\end{figure}
Efficient and precise preparation of random quantum states constitutes a critical requirement for the practical implementation of the RH architecture.
There are many approaches can be used to generate random quantum states. Such as the circuit whose gates are sampled and deployed randomly \cite{random_circuits_sampling}, and the amplitude encoding circuit \cite{amplitude_encoding} whose parameters are randomly assigned.
Nevertheless, these circuits are quite deep, especially the latter, which gives rise of exponential experimental cost.
Here we devise a \textit{local random circuit}, which is composed of a series of single-qubit
rotation gates as illustrated in Figure~\ref{fig:11} and generates
quantum states that are the tensor product of $n$ single-qubit random
states from the initial state $|0\rangle^{n}$. The amplitude of the resultant local random state on the $i^{\text{th}}$ computational basis is
\begin{equation}
	|\varphi\rangle_i=\prod\nolimits_{k=0}^{n-1}|\varphi_k\rangle_{\text{bin}(i)_k},
\end{equation}
where $|\varphi_k\rangle$ is the single-qubit random state encoded in the $k^{\text{th}}$ qubit, and the function $\text{bin}(\cdot)$ outputs the binary representation of its input.
The amplitudes are mutually independent (ignoring a trivial global normalization factor) as they differ by at least one random number, thereby fulfilling the prerequisites for applying Theorem~\ref{theorem1}.

It worth stating that
although this circuit cannot generate truly entangled states, they
will \textit{arithmetically} exist as components in a certain amount in the resultant
states. For example, $(|00\rangle+|01\rangle+|10\rangle+|11\rangle)/2$
can be viewed arithmetically as a superposition of entangled states
$(|00\rangle+|11\rangle)/\sqrt{2}$ and $(|01\rangle+|10\rangle)/\sqrt{2}$.

\section{Experiments}\label{section_experiment}
Many important applications are naturally categorized as quantum circuit redesign problems. For the sake of concreteness, we apply our RH architecture to three applications: non-parameterized (and parameterized) quantum circuit equivalence checking, and quantum circuit variational reconstruction. These applications capture the core of quantum circuit redesign and simultaneously are limited to a small-scale (4-qubit) to be performable by current quantum devices. Larger-scale validation experiments ($4\sim9$ qubits) based purely on classical simulation are presented in Appendix~\ref{appendices_b}.
\begin{figure}[h] 
	\centering
	\subfigure[ ]{\includegraphics[width=0.4\linewidth]{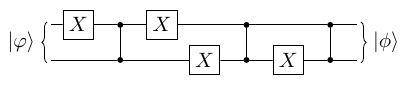}}\subfigure[ ]{\includegraphics[width=0.36\linewidth]{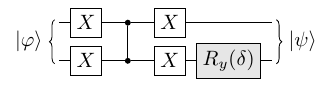}}\subfigure[ ]{\includegraphics[width=0.24\linewidth]{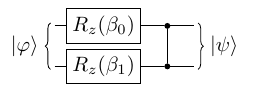}}\\
	\subfigure[ ]{\includegraphics[width=0.48\linewidth]{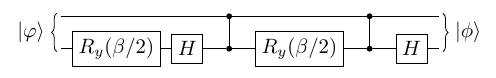}}\subfigure[ ]{\includegraphics[width=0.52\linewidth]{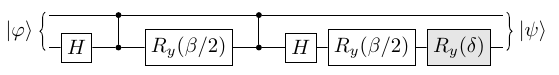}}
	\caption{\label{fig_applications}
		(a) A two-qubit phase-flipper circuit for basis states $|01\rangle$, $|10\rangle$ and $|11\rangle$. 
		(b) A two-qubit phase-flipper circuit for basis state $|00\rangle$.
		(c) A potential alternative to the circuit (a) implemented by one two-qubit $CZ$ gate and two single-qubit rotation gates.
		(d) and (e)	are two decomposition circuits which are frequently
		employed for implementing the $CR_{y}(\theta)$ gate. 
		The gray $R_{y}(\delta)$ gates in circuits (b) and (e) are introduced to create a controlled amount of distortion relative to their original functionality.}
\end{figure}
To assess the performance of the RH architecture, we conduct these applications both in numerical simulation and in current NISQ hardware devices. The numerical simulation is performed with Python 3.10, MindSpore Quantum 0.10 \cite{MindQuantum} and have been run on an 4-core 1.8 GHz CPU with 8 GB memory. The hardware experiments are carried out on the Zuchongzhi-$2$ quantum processor, a $66$-qubit superconducting quantum platform that was used to demonstrate quantum advantage \cite{zuchongzhi21}. Detailed hardware experiment information, including error rates and qubit-mapping schematic, is provided in Appendix~\ref{appendices_a}. The code and data used in this paper are publicly available on Gitee at \url{https://gitee.com/mindspore/mindquantum/tree/research/paper_with_code}.
\begin{figure}[h] 
	\centering
	\subfigure[ ]{\includegraphics[width=0.45\linewidth]{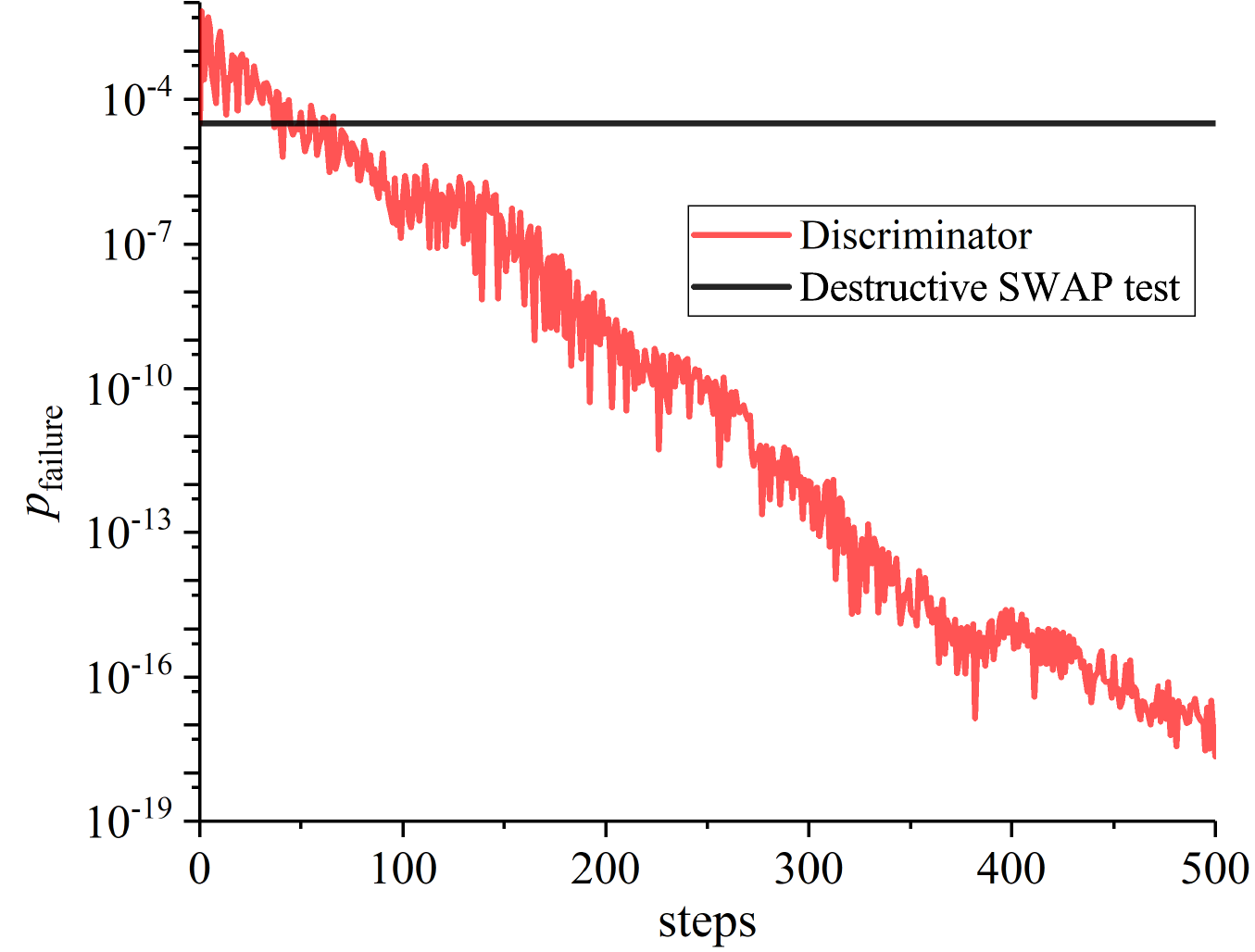}}\subfigure[ ]{\includegraphics[width=0.45\linewidth]{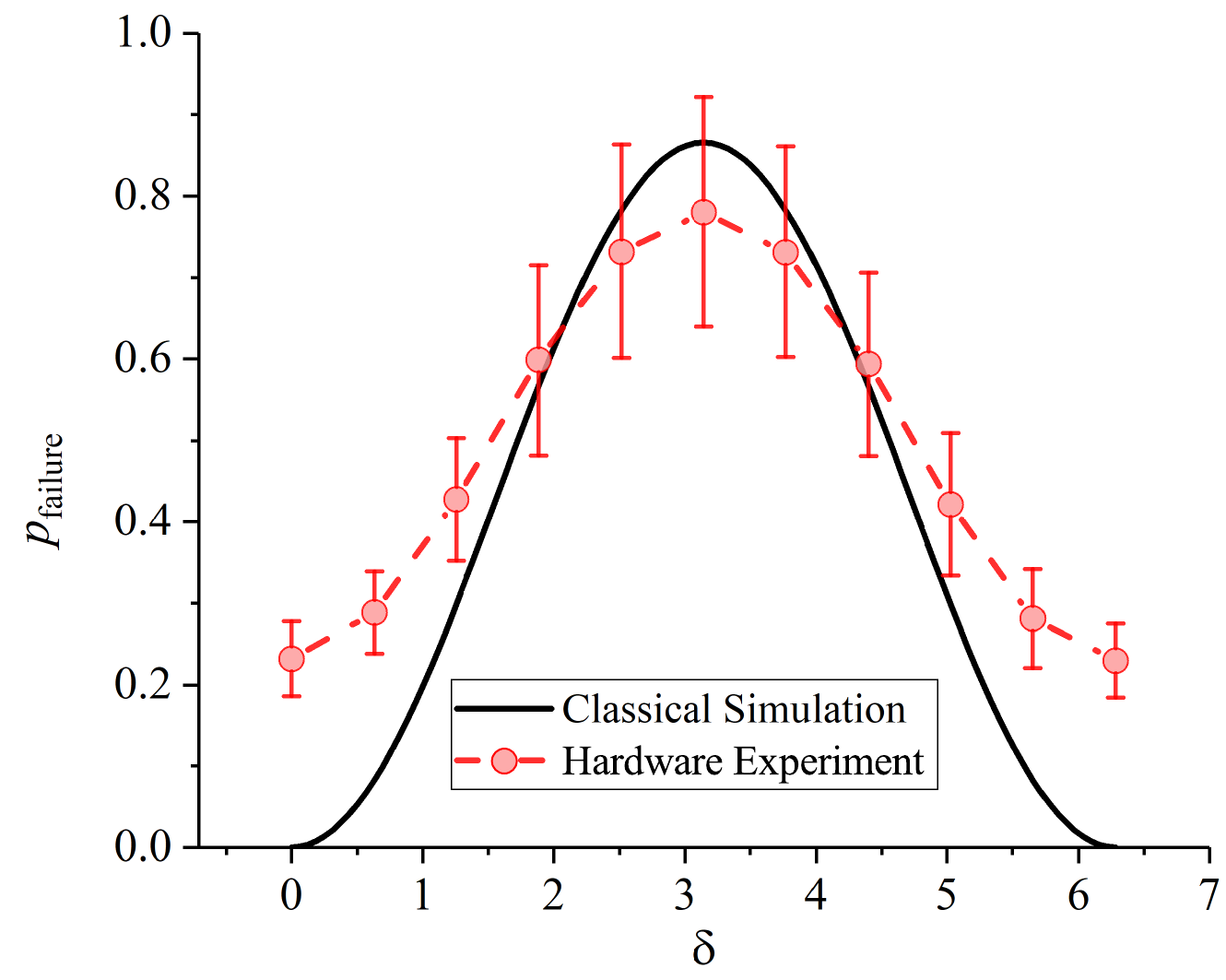}}
	\subfigure[ ]{\includegraphics[width=0.45\linewidth]{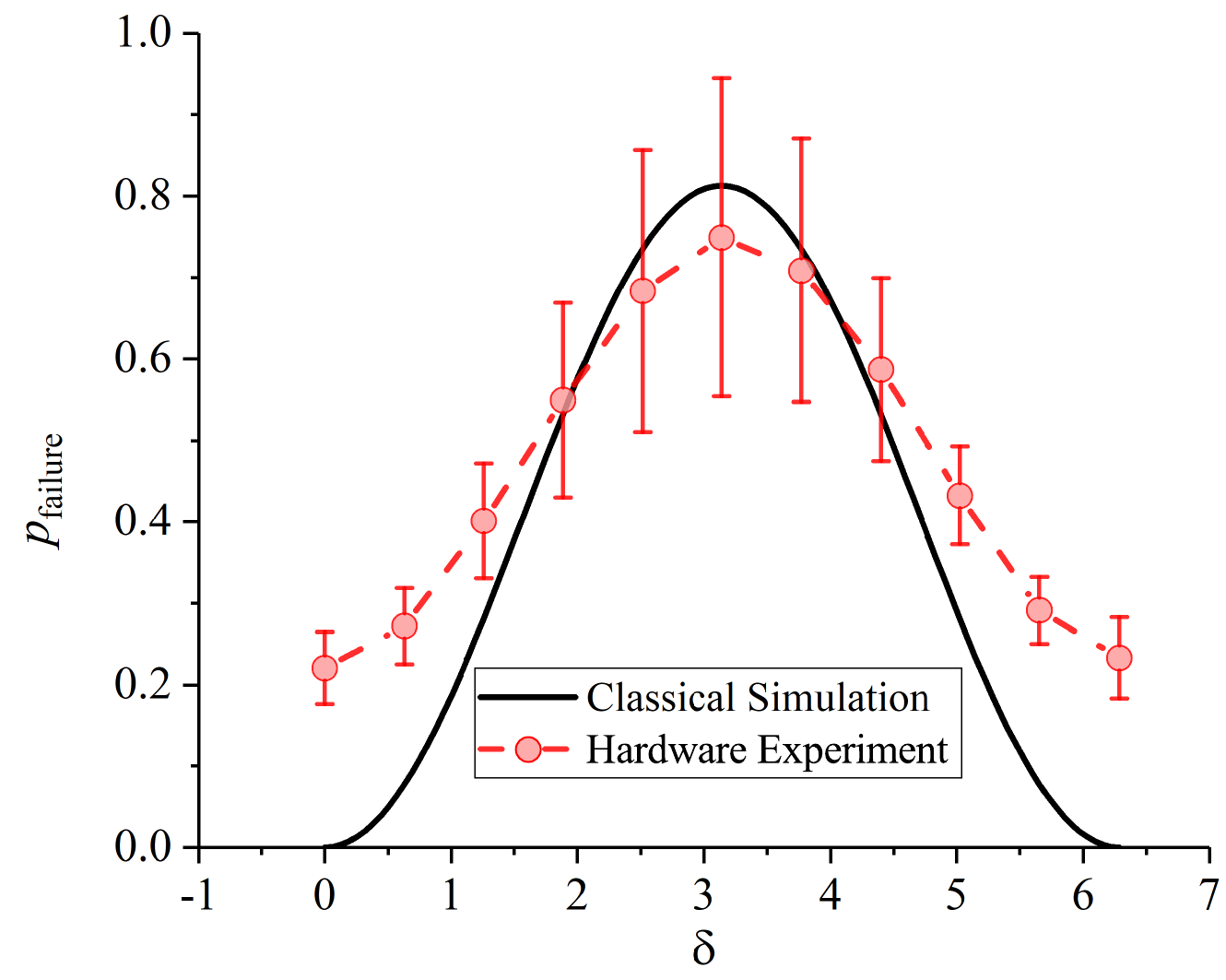}}\subfigure[ ]{\includegraphics[width=0.45\linewidth]{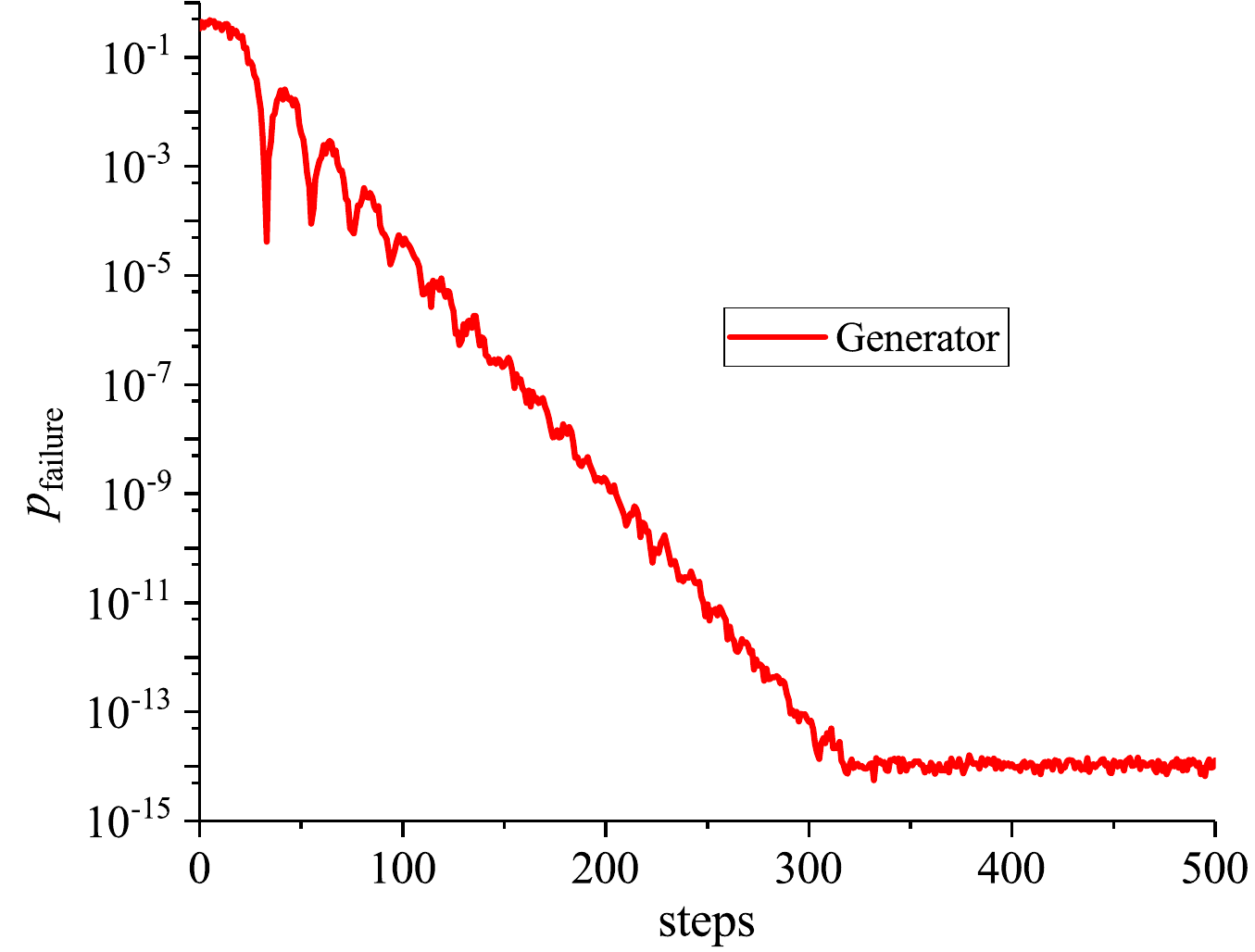}}
	\caption{\label{fig_results} 
		(a) The probabilities of test failure as functions of steps in training the discriminator classically.
		(b) The averaged probabilities of test failure (with $\pm1$ standard deviation error bars) vary with the value of parameter $\delta$ in the circuit of Figure~\ref{fig_applications} (b).
		(c) The averaged probabilities of test failure  (with $\pm1$ standard deviation error bars) exhibit a parametric dependence on $\delta$ in the circuit of Figure~\ref{fig_applications} (e).
		(d) The averaged probability of test failure evolves with respect to the training steps in classical simulation.}
\end{figure}
\subsection{Preliminary Experiment}
Before implementing these applications, we first train the discriminator circuit -- a 4-qubit parameterized destructive SWAP test circuit \cite{destructive_swap_test} (Figure~\ref{fig_eq_gan} (b)). 
For training the discriminator efficiently, we feed it two zero states, i.e. $|\psi\rangle=|\phi\rangle=|\varphi\rangle=|0\rangle^{\otimes2}$ with reasons explained in Subsection~\ref{workflow}, and optimize its parameters $\boldsymbol{\theta}$ to minimize the $p_{\text{failure}}$, allowing it to eventually act as an effective destructive SWAP test circuit in a noisy environment.
The well trained discriminator obtained here will be reused in following applications.


To evaluate the discriminator's robustness against errors, we introduce two additional $R_{z}$ gates (not shown in the figure) after each CZ gate in the classical simulation, modeling the effect of coherent errors \cite{QGAN_2}.
The parameters of these  $R_{z}$ gates are randomly sampled from a normal distribution with $\mu=0$ and $\sigma=0.02$. The discriminator's trainable parameters $\boldsymbol{\theta}$ are initialized to zero and optimized using the Adam optimizer with an initial learning rate of $0.1$. When all parameters  $\boldsymbol{\theta}=\boldsymbol{0}$, the parameterized discriminator  degenerated into the original destructive SWAP test. 

In Figure~\ref{fig_results} (a), we depict the evolution of the cost function $p_{\text{failure}}$ with respect to the training steps in training the discriminator classically. We can see that within $500$ training
steps the probability of test failure gradually decreases, from the
initial value of $3.148\times10^{-4}$ to a minimum of $2.232\times10^{-17}$.
In hardware experiment, the $p_{\text{failure}}$ decreases
from the initial $9.400\times10^{-2}$ to a minimum of $5.150\times10^{-2}$ within $50$ training steps (without presenting in figure). The number of shots in experiment is $1000$ throughout this paper.

\begin{table*}
	\caption{The comparison of the discriminator's performance before
		and after training, along with the values of its parameters. The notation
		$\boldsymbol{\theta}$ refers to the results from classical simulation, while $\boldsymbol{\theta}^{*}$ refers to the results from hardware experiment. }
	\label{tab:1} 
	\begin{tabular}{cccccc}
		\toprule
		$p_{\text{failure}}$ & \diagbox{}{$i$}  & 0 & 1 & 2 & 3\\
		\hline 
		$1.291\times10^{-3}$ & $\boldsymbol{\theta}$ & 0.0 & 0.0 & 0.0 & 0.0\\
		$1.214\times10^{-9}$ & $\boldsymbol{\theta}$ & $4.981\times10^{-3}$ & $-1.088\times10^{-2}$ & $6.910\times10^{-3}$ & $-8.094\times10^{-3}$\\
		$9.104\times10^{-2}$ & $\boldsymbol{\theta}^{*}$ & 0.0 & 0.0 & 0.0 & 0.0\\
		$7.862\times10^{-2}$ & $\boldsymbol{\theta}^{*}$ & $1.692\times10^{-2}$ & $1.618\times10^{-2}$ & $-1.114\times10^{-2}$ & $5.916\times10^{-2}$\\
		--- & error & $-4.981\times10^{-3}$ & $1.088\times10^{-2}$ & $-6.910\times10^{-3}$ & $8.093\times10^{-3}$\\
	\bottomrule
	\end{tabular}
\end{table*}

In Table~\ref{tab:1}, we record the performance comparison of the discriminator before and after training, as well as the values of its parameters. Referring to Table~\ref{tab:1}, after training, not only the discriminator reduces the probability of test failure from $1.291\times10^{-3}$
to $1.214\times10^{-9}$ in simulation, but also the optimized parameters perfectly compensate for the randomly instantiated coherent error which is listed in the last row, i.e. $\theta_{i}=-\text{noise}_{i}$, achieving perfect error cancellation. These results indicate that the discriminator actually acquires remarkable error robustness by training.
We emphasize here that during the training process, the $p_{\text{failure}}$
is calculated for zero states, while in the final performance evaluation
in Table~\ref{tab:1}, the $p_{\text{failure}}$ is averaged over $100$
random quantum states. Their specific values might vary, yet the main
conclusion holds.

\subsection{Application 1: Non-parameterized Quantum Circuit Equivalence Checking}
The first application to be outlined is checking the equivalence between
two non-parameterized circuits. We take the circuits
shown in Figure~\ref{fig_applications} (a) and (b) as the reference circuit and
the generator circuit in RH architecture respectively. Ignoring
the $R_{y}(\delta)$ gate in gray, these two circuits are phase-flippers and have important applications in oracle-based quantum algorithms, such as the Grover's searching algorithm \cite{grover_0}.  
When $\delta=2k\pi$ ($k$ is an integer), these two structurally distinct circuits are equivalent up to a global phase $-1$, and thus may generated by different synthesis tools for the same unitary matrix. The necessity of carrying out an equivalence checking on them naturally arises. To achieve this, we take $100$ random states $|\varphi\rangle$ as inputs and study the averaged probability of test failure as the parameter $\delta$ of the $R_{y}$ gate in gray alters.

From Figure~\ref{fig_results} (b), we see that as $\delta$ increases within the interval $[0,2\pi]$, the probability of test failure first rises from $0$ and then falls back, which precisely conforms to the theoretical expectations. Similar conclusion can also be drawn according to the hardware results, albeit with deviations caused by noise and operational imperfections.

\subsection{Application 2: Parameterized Circuit Equivalence Checking}
Owing to the current prevalence of quantum machine learning, checking
the equivalence of parameterized circuits is a highly important but
simultaneously intractable problem, and there is a lack of universal
and effective solutions \cite{ZX_for_circuit_equivalence}. 
In this Subsection, as a representative example, we consider the
equivalence checking between two decomposition circuits of the $CR_{y}(\beta)$
gate, which are shown in Figure~\ref{fig_applications} (d) and (e), respectively.
These two parameterized circuits are equivalent for any real $\beta$ when
the parameter of the $R_{y}(\delta)$  gate in gray satisfies $\delta=2k\pi$, where
$k$ is an integer, and deviates from each other in other cases. 

Figure~\ref{fig_results} (c) describes the averaged $p_{\text{failure}}$ in classical simulation and hardware experiment over $10$ random
states and $10$ random $\beta$ in terms of the value of $\delta$ in
the interval $[0,2\pi]$. The feasibility of checking equivalence
between parameterized circuits with random free parameters, i.e.,
the $\beta$ herein, has been proven in Reference~\cite{ZX_for_circuit_equivalence}.
It can be concluded from Figure~\ref{fig_results} (c) that the probabilities
of test failure nicely reflect the amount of deviation between the
two circuits as the $\delta$ varies. 

\subsection{Application 3: Quantum Circuit Variational Optimization}
Circuit optimization seeks to construct equivalent quantum circuits with either reduced gate counts or shallower depths compared to their target circuits. For complex circuit optimization, variational algorithms often provide more reliable and systematic approaches than developer intuition alone \cite{he_MS}. 

The example we present here is to reduce the depth of a particular circuit illustrated in Figure~\ref{fig_applications} (a) in a variational manner.  A potential alternate is depicted in Figure~\ref{fig_applications} (c), which comprises two single-qubit $R_{z}$ gates with trainable parameters  $\beta_{0}$ and $\beta_{1}$, as well as one $CZ$  gate.
Within the RH architecture, we designate these two circuits as the
reference circuit and the generator circuit respectively. In this
example, the settings of the hyper-parameters are aligned with those
in the previous training of the discriminator, such as the parameters
initialization, learning rate, cost function and so on. 

Figure~\ref{fig_results} (d) reveals that
as the training steps increasing, the $p_{\text{failure}}$ in classical
simulation steadily decreases from an initial value of $6.654\times10^{-1}$ and
converges to around $1\times10^{-14}$ after about $300$ steps. The
results from hardware experiment show that the probability of test
failure has declined from the initial value of $8.66\times10^{-1}$
to a minimum of $1.099\times10^{-1}$ during a 50-step training procedure
(without presenting in figure). The $p_{\text{failure}}$ 
during training is dynamically computed using a batch of $4$ random states. These observations imply that the RH architecture works well in this application, and the resultant generator circuit can be a competent alternative to the reference circuit in experiment after training.

%

\section{Conclusion}


In this paper, we formally defined the task of quantum circuit redesign, and introduced a novel RH architecture for redesigning large-size quantum circuits on quantum hardware that overcomes the classical simulation bottleneck inherent to existing methods.
The RH architecture is built upon the EQ-GAN, and extends its capability  from quantum states learning to unitary transformation learning by inserting a quantum random circuit module prior to it. We proposed the usage of $|0\rangle^n$ state in training the discriminator to minimize resource overhead and error accumulation, and demonstrated the discriminator's robustness against coherent errors of $CZ$ gate on superconducting quantum processors.
We proved the RH architecture's completeness and discussed its limitations. An efficient local random quantum circuit was developed that drastically slashes implementation overhead, boosting practical utility.
As demonstrations, we implemented the RH architecture to three critical applications: (non) parameterized circuit equivalence checking and variational reconstruction of quantum circuits.
Evaluation results from both classical simulation and quantum hardware experiments confirm the superior performance of the RH architecture.

The RH architecture is also compatible with other existing circuit optimization techniques like ZX-calculus, enabling collaborative solutions for advanced circuit optimization challenges.
We believe our RH architecture could help boost innovation in
designing advanced quantum circuits.

\bibliographystyle{plainnat}  
\bibliography{TempReferences}

\begin{thebibliography}{34}
\providecommand{\natexlab}[1]{#1}
\providecommand{\url}[1]{\texttt{#1}}
\expandafter\ifx\csname urlstyle\endcsname\relax
  \providecommand{\doi}[1]{doi: #1}\else
  \providecommand{\doi}{doi: \begingroup \urlstyle{rm}\Url}\fi

\bibitem[Arute et~al.(2019)Arute, Arya, Babbush, Bacon, Bardin, Barends, Biswas, Boixo, Brandao, Buell, et~al.]{google_2019}
Frank Arute, Kunal Arya, Ryan Babbush, Dave Bacon, Joseph~C Bardin, Rami Barends, Rupak Biswas, Sergio Boixo, Fernando~GSL Brandao, David~A Buell, et~al.
\newblock Quantum supremacy using a programmable superconducting processor.
\newblock \emph{Nature}, 574\penalty0 (7779):\penalty0 505--510, 2019.
\newblock URL \url{https://doi.org/10.1038/s41586-019-1666-5}.

\bibitem[Barbeau and Garcia-Alfaro(2019)]{QGAN_0}
Michel Barbeau and Joaquin Garcia-Alfaro.
\newblock Faking and discriminating the navigation data of a micro aerial vehicle using quantum generative adversarial networks.
\newblock In \emph{2019 IEEE Globecom Workshops (GC Wkshps)}, pages 1--6. IEEE, 2019.

\bibitem[Benedetti et~al.(2019)Benedetti, Lloyd, Sack, and Fiorentini]{QML_review_QST}
Marcello Benedetti, Erika Lloyd, Stefan Sack, and Mattia Fiorentini.
\newblock Parameterized quantum circuits as machine learning models.
\newblock \emph{Quantum Science and Technology}, 4\penalty0 (4):\penalty0 043001, 2019.

\bibitem[Bishnoi(2020)]{qc_bb}
Bhupesh Bishnoi.
\newblock Quantum computation.
\newblock \emph{arXiv preprint arXiv:2006.02799}, 2020.
\newblock URL \url{https://arxiv.org/abs/2006.02799}.

\bibitem[Boixo et~al.(2018)Boixo, Isakov, Smelyanskiy, Babbush, Ding, Jiang, Bremner, Martinis, and Neven]{random_circuits_sampling}
Sergio Boixo, Sergei~V Isakov, Vadim~N Smelyanskiy, Ryan Babbush, Nan Ding, Zhang Jiang, Michael~J Bremner, John~M Martinis, and Hartmut Neven.
\newblock Characterizing quantum supremacy in near-term devices.
\newblock \emph{Nature Physics}, 14\penalty0 (6):\penalty0 595--600, 2018.
\newblock URL \url{https://doi.org/10.1038/s41567-018-0124-x}.

\bibitem[Buhrman et~al.(2001)Buhrman, Cleve, Watrous, and De~Wolf]{swap_test}
Harry Buhrman, Richard Cleve, John Watrous, and Ronald De~Wolf.
\newblock Quantum fingerprinting.
\newblock \emph{Physical review letters}, 87\penalty0 (16):\penalty0 167902, 2001.

\bibitem[Burgholzer and Wille(2020)]{Decision_diagram_for_circuit_equivalence}
Lukas Burgholzer and Robert Wille.
\newblock Advanced equivalence checking for quantum circuits.
\newblock \emph{IEEE Transactions on Computer-Aided Design of Integrated Circuits and Systems}, 40\penalty0 (9):\penalty0 1810--1824, 2020.

\bibitem[Chen et~al.(2022)Chen, Fang, Guan, Hong, Huang, Liu, Wang, and Ying]{veriqbench}
Kean Chen, Wang Fang, Ji~Guan, Xin Hong, Mingyu Huang, Junyi Liu, Qisheng Wang, and Mingsheng Ying.
\newblock Veriqbench: A benchmark for multiple types of quantum circuits, 2022.
\newblock URL \url{https://arxiv.org/abs/2206.10880}.

\bibitem[Dawson and Nielsen(2005)]{SK_algorithm}
Christopher~M Dawson and Michael~A Nielsen.
\newblock The solovay-kitaev algorithm.
\newblock \emph{arXiv preprint quant-ph/0505030}, 2005.
\newblock URL \url{https://arxiv.org/abs/quant-ph/0505030}.

\bibitem[Ding et~al.(2006)Ding, Jin, and Yang]{evolutionary_algorithm_for_oracle}
Shengchao Ding, Zhi Jin, and Qing Yang.
\newblock Evolving quantum oracles with hybrid quantum-inspired evolutionary algorithm, 2006.
\newblock URL \url{https://arxiv.org/abs/quant-ph/0610105}.

\bibitem[Garcia-Escartin and Chamorro-Posada(2013)]{destructive_swap_test}
Juan~Carlos Garcia-Escartin and Pedro Chamorro-Posada.
\newblock Swap test and hong-ou-mandel effect are equivalent.
\newblock \emph{Physical Review A}, 87\penalty0 (5):\penalty0 052330, 2013.

\bibitem[Ge et~al.(2024)Ge, Wenjie, Yuheng, Kaisen, Xudong, Zixiang, Yuhan, Ruocheng, and Junchi]{circuit_synthesis_review}
Yan Ge, Wu~Wenjie, Chen Yuheng, Pan Kaisen, Lu~Xudong, Zhou Zixiang, Wang Yuhan, Wang Ruocheng, and Yan Junchi.
\newblock Quantum circuit synthesis and compilation optimization: Overview and prospects.
\newblock \emph{arXiv preprint arXiv:2407.00736}, 2024.

\bibitem[Goodfellow et~al.(2016)Goodfellow, Bengio, and Courville]{deep_learning_book}
Ian Goodfellow, Yoshua Bengio, and Aaron Courville.
\newblock \emph{Deep Learning}.
\newblock The MIT Press, USA, 2016.
\newblock ISBN 0262035618.

\bibitem[Grover(1996)]{grover_0}
Lov~K Grover.
\newblock A fast quantum mechanical algorithm for database search.
\newblock In \emph{Proceedings of the twenty-eighth annual ACM symposium on Theory of computing}, pages 212--219, 1996.

\bibitem[Harrow et~al.(2009)Harrow, Hassidim, and Lloyd]{hhl}
Aram~W Harrow, Avinatan Hassidim, and Seth Lloyd.
\newblock Quantum algorithm for linear systems of equations.
\newblock \emph{Physical review letters}, 103\penalty0 (15):\penalty0 150502, 2009.

\bibitem[He et~al.(2021)He, Wang, Wu, Nie, Zhang, and Wang]{he_DRL}
Run-Hong He, Rui Wang, Jing Wu, Shen-Shuang Nie, Jia-Hui Zhang, and Zhao-Ming Wang.
\newblock Deep reinforcement learning for universal quantum state preparation via dynamic pulse control, 2021.
\newblock URL \url{https://doi.org/10.1140/epjqt/s40507-021-00119-6}.

\bibitem[He et~al.(2023)He, Xu, Byrd, and Wang]{he_MS}
Run-Hong He, Xu-Sheng Xu, Mark~S Byrd, and Zhao-Ming Wang.
\newblock Modularized and scalable compilation for double quantum dot quantum computing.
\newblock \emph{Quantum Science and Technology}, 9\penalty0 (1):\penalty0 015004, oct 2023.
\newblock \doi{10.1088/2058-9565/acfe38}.
\newblock URL \url{https://dx.doi.org/10.1088/2058-9565/acfe38}.

\bibitem[Hu et~al.(2019)Hu, Wu, Cai, Ma, Mu, Xu, Wang, Song, Deng, Zou, et~al.]{QGAN_in_superconding_system}
Ling Hu, Shu-Hao Wu, Weizhou Cai, Yuwei Ma, Xianghao Mu, Yuan Xu, Haiyan Wang, Yipu Song, Dong-Ling Deng, Chang-Ling Zou, et~al.
\newblock Quantum generative adversarial learning in a superconducting quantum circuit.
\newblock \emph{Science advances}, 5\penalty0 (1):\penalty0 eaav2761, 2019.

\bibitem[Long(2001)]{grover_3}
Gui-Lu Long.
\newblock Grover algorithm with zero theoretical failure rate.
\newblock \emph{Physical Review A}, 64\penalty0 (2):\penalty0 022307, 2001.

\bibitem[M{\"o}tt{\"o}nen et~al.(2004)M{\"o}tt{\"o}nen, Vartiainen, Bergholm, and Salomaa]{cs_decompose}
Mikko M{\"o}tt{\"o}nen, Juha~J Vartiainen, Ville Bergholm, and Martti~M Salomaa.
\newblock Quantum circuits for general multiqubit gates.
\newblock \emph{Physical review letters}, 93\penalty0 (13):\penalty0 130502, 2004.

\bibitem[Nielsen and Chuang(2010)]{qc_nielsen}
Michael~A Nielsen and Isaac~L Chuang.
\newblock \emph{Quantum Computation and Quantum Information 10th Anniversary Edition}.
\newblock Cambridge University Press, 2010.

\bibitem[Niu et~al.(2022)Niu, Zlokapa, Broughton, Boixo, Mohseni, Smelyanskyi, and Neven]{QGAN_2}
Murphy~Yuezhen Niu, Alexander Zlokapa, Michael Broughton, Sergio Boixo, Masoud Mohseni, Vadim Smelyanskyi, and Hartmut Neven.
\newblock Entangling quantum generative adversarial networks.
\newblock \emph{Physical Review Letters}, 128\penalty0 (22):\penalty0 220505, 2022.

\bibitem[Peham et~al.(2023)Peham, Burgholzer, and Wille]{ZX_for_circuit_equivalence}
Tom Peham, Lukas Burgholzer, and Robert Wille.
\newblock Equivalence checking of parameterized quantum circuits: Verifying the compilation of variational quantum algorithms.
\newblock In \emph{Proceedings of the 28th Asia and South Pacific Design Automation Conference}, pages 702--708, 2023.

\bibitem[Quantum et~al.(2020)Quantum, Collaborators*†, Arute, Arya, Babbush, Bacon, Bardin, Barends, Boixo, Broughton, Buckley, et~al.]{rz_for_cz_gate}
Google~AI Quantum, Collaborators*†, Frank Arute, Kunal Arya, Ryan Babbush, Dave Bacon, Joseph~C Bardin, Rami Barends, Sergio Boixo, Michael Broughton, Bob~B Buckley, et~al.
\newblock Hartree-fock on a superconducting qubit quantum computer.
\newblock \emph{Science}, 369\penalty0 (6507):\penalty0 1084--1089, 2020.

\bibitem[Rakyta and Zimbor{\'a}s(2022{\natexlab{a}})]{gate_decompose_with_VQA}
P{\'e}ter Rakyta and Zolt{\'a}n Zimbor{\'a}s.
\newblock Approaching the theoretical limit in quantum gate decomposition.
\newblock \emph{Quantum}, 6:\penalty0 710, 2022{\natexlab{a}}.

\bibitem[Rakyta and Zimbor{\'a}s(2022{\natexlab{b}})]{gate_decompose_with_VQA_adaptive_circuit_compression}
P{\'e}ter Rakyta and Zolt{\'a}n Zimbor{\'a}s.
\newblock Efficient quantum gate decomposition via adaptive circuit compression.
\newblock \emph{arXiv preprint arXiv:2203.04426}, 2022{\natexlab{b}}.

\bibitem[Rasmussen and Zinner(2022)]{QGAN_1}
SE~Rasmussen and NT~Zinner.
\newblock Multiqubit state learning with entangling quantum generative adversarial networks.
\newblock \emph{Physical Review A}, 106\penalty0 (3):\penalty0 032429, 2022.

\bibitem[Shende et~al.(2005)Shende, Bullock, and Markov]{amplitude_encoding}
Vivek~V Shende, Stephen~S Bullock, and Igor~L Markov.
\newblock Synthesis of quantum logic circuits.
\newblock In \emph{Proceedings of the 2005 Asia and South Pacific Design Automation Conference}, pages 272--275, 2005.

\bibitem[Shor(1994)]{shor}
Peter~W Shor.
\newblock Algorithms for quantum computation: discrete logarithms and factoring.
\newblock In \emph{Proceedings 35th annual symposium on foundations of computer science}, pages 124--134. Ieee, 1994.

\bibitem[Wang et~al.(2020)Wang, Wang, Li, Fan, Wei, and Gu]{wangshengbin_hhl_poisson_equations}
Shengbin Wang, Zhimin Wang, Wendong Li, Lixin Fan, Zhiqiang Wei, and Yongjian Gu.
\newblock Quantum fast poisson solver: the algorithm and complete and modular circuit design.
\newblock \emph{Quantum Information Processing}, 19:\penalty0 1--25, 2020.

\bibitem[Williams and Gray(1998)]{genetic_algorithm_for_logic_circuit_synthesis}
Colin~P Williams and Alexander~G Gray.
\newblock Automated design of quantum circuits.
\newblock In \emph{NASA International Conference on Quantum Computing and Quantum Communications}, pages 113--125. Springer, 1998.

\bibitem[Xu et~al.(2024)Xu, Cui, Cui, He, Li, Li, Lin, Liu, Liu, Lu, Luo, Lyu, Pan, Pavel, Shu, Tang, Xu, Xu, Yang, Yu, Zeng, Zhao, Zheng, Zhou, Zhou, Zhu, Zou, Bayat, Cao, Cui, Li, Long, Su, Wang, Wang, Wei, Wu, Zhang, and Yung]{MindQuantum}
Xusheng Xu, Jiangyu Cui, Zidong Cui, Runhong He, Qingyu Li, Xiaowei Li, Yanling Lin, Jiale Liu, Wuxin Liu, Jiale Lu, Maolin Luo, Chufan Lyu, Shijie Pan, Mosharev Pavel, Runqiu Shu, Jialiang Tang, Ruoqian Xu, Shu Xu, Kang Yang, Fan Yu, Qingguo Zeng, Haiying Zhao, Qiang Zheng, Junyuan Zhou, Xu~Zhou, Yikang Zhu, Zuoheng Zou, Abolfazl Bayat, Xi~Cao, Wei Cui, Zhendong Li, Guilu Long, Zhaofeng Su, Xiaoting Wang, Zizhu Wang, Shijie Wei, Re-Bing Wu, Pan Zhang, and Man-Hong Yung.
\newblock Mindspore quantum: A user-friendly, high-performance, and ai-compatible quantum computing framework, 2024.
\newblock URL \url{https://arxiv.org/abs/2406.17248}.

\bibitem[Zhang et~al.(2020)Zhang, Zheng, Zhang, and Deng]{gate_decompose_prl_dqn}
Yuan-Hang Zhang, Pei-Lin Zheng, Yi~Zhang, and Dong-Ling Deng.
\newblock Topological quantum compiling with reinforcement learning.
\newblock \emph{Physical Review Letters}, 125\penalty0 (17):\penalty0 170501, 2020.

\bibitem[Zhu et~al.(2022)Zhu, Cao, Chen, Chen, Chen, Chung, Deng, Du, Fan, Gong, Guo, Guo, Guo, Han, Hong, Huang, Huo, Li, Li, Li, Li, Liang, Lin, Lin, Qian, Qiao, Rong, Su, Sun, Wang, Wang, Wu, Wu, Xu, Yan, Yang, Yang, Ye, Yin, Ying, Yu, Zha, Zhang, Zhang, Zhang, Zhang, Zhao, Zhao, Zhou, Lu, Peng, Zhu, and Pan]{zuchongzhi21}
Qingling Zhu, Sirui Cao, Fusheng Chen, Ming-Cheng Chen, Xiawei Chen, Tung-Hsun Chung, Hui Deng, Yajie Du, Daojin Fan, Ming Gong, Cheng Guo, Chu Guo, Shaojun Guo, Lianchen Han, Linyin Hong, He-Liang Huang, Yong-Heng Huo, Liping Li, Na~Li, Shaowei Li, Yuan Li, Futian Liang, Chun Lin, Jin Lin, Haoran Qian, Dan Qiao, Hao Rong, Hong Su, Lihua Sun, Liangyuan Wang, Shiyu Wang, Dachao Wu, Yulin Wu, Yu~Xu, Kai Yan, Weifeng Yang, Yang Yang, Yangsen Ye, Jianghan Yin, Chong Ying, Jiale Yu, Chen Zha, Cha Zhang, Haibin Zhang, Kaili Zhang, Yiming Zhang, Han Zhao, Youwei Zhao, Liang Zhou, Chao-Yang Lu, Cheng-Zhi Peng, Xiaobo Zhu, and Jian-Wei Pan.
\newblock Quantum computational advantage via 60-qubit 24-cycle random circuit sampling.
\newblock \emph{Science Bulletin}, 67\penalty0 (3):\penalty0 240--245, 2022.
\newblock ISSN 2095-9273.
\newblock \doi{https://doi.org/10.1016/j.scib.2021.10.017}.
\newblock URL \url{https://www.sciencedirect.com/science/article/pii/S2095927321006733}.

\end{thebibliography}

\newpage
\appendix
\section{Technical Appendices and Supplementary Material}
\subsection{Hardware Information \label{appendices_a}}

\begin{figure}[h] 
	\centering
	\includegraphics[scale=0.5]{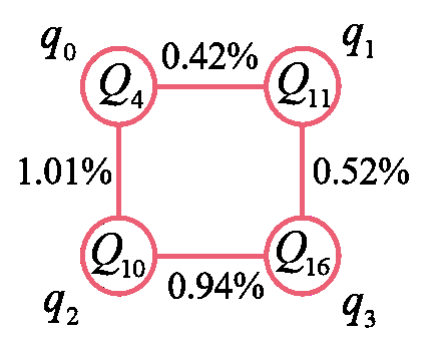}
	\caption{\label{fig_qubit_mapping} The mapping schematic between the logical
	qubits and the physical qubits arranged in a $2\times 2$ sub-grid within
	the Zuchongzhi-$2$ quantum processor. The hardware was calibrated on
	December $11$th, $2024$.}
\end{figure}
To access results from quantum hardware, we map the utilized $4$ logical
qubits $q_{i}$ onto physical qubits $Q_{i}$ in a $2\times2$ sub-grid
of the 66-qubit Zuchongzhi-$2$ quantum processor for accommodating
the experimental realities. The mapping diagram is illustrated in
Figure~\ref{fig_qubit_mapping} (c), where the label of line between physical qubits indicates the error rate in executing $CZ$ gate. The error rates
of single-qubit native gates are between $0.11\%$ and $0.21\%$, and the average readout error rate is $4.21\%$.

\subsection{Advanced Experiments \label{appendices_b}}
In Section~\ref{section_experiment}, we demonstrated the workflow of our RH architecture through three small-scale proof-of-concept experiments.
We now consider larger-scale scenarios. Since quantum circuit equivalence verification constitutes the core contribution of the RH framework and serves as the foundation for all its applications, this section focuses on benchmarking its performance for various commonly used quantum circuits. The benchmark circuits are selected from the VeriQBench dataset \cite{veriqbench}, covering major quantum computing applications, such as the quantum machine learning, variational quantum computational chemistry, Grover's algorithm, quantum fourier transform, qubit mapping and quantum volume tests, etc. 

Given current quantum hardware's noise limitations for deep multi-qubit circuits,  we restrict our benchmarking to only classical simulations.  The tested circuits are limited to $n = 4\sim9$ qubits (RH architecture: $2n = 8\sim18$ qubits) to comply with the computational constraints of classical simulators. We stress that the RH architecture is theoretically scalable to arbitrary circuit sizes. With anticipated advancements in quantum hardware, particularly fault-tolerant quantum computers, verification of circuits comprising hundreds to thousands of qubits will become experimentally feasible.

To evaluate RH's performance, we prepare generator circuits by inserting quantum gates at random positions in given reference circuits. 
We generate equivalent circuit by inserting an identity gate into the given reference circuit, while creating non-equivalent circuit through the insertion of $R_x(1.23)$ rotation gate. As for the parameterized circuits, such as the `qaoa\_6' and `qcnn\_4', we randomly instantiate their parameters, as justified by the theoretical proof in Reference~\cite{ZX_for_circuit_equivalence}.

The benchmark results are summarized in Table~\ref{tab_benchmarks}, where $p_{\text{failure\_Y}}$ denotes the test failure probability for equivalent circuit pairs and $p_{\text{failure\_N}}$ represents the failure probability for non-equivalent pairs under the RH architecture, respectively.

As evidenced by the benchmark data in Table~\ref{tab_benchmarks}, the test failure probability $p_{\text{failure\_Y}}$ remains zero within machine precision ($\sim 10^{-31}$) for equivalent circuit pairs. In contrast, for non-equivalent pairs, the test failure probability $p_{\text{failure\_N}}$ ranges from $0.155$ to $0.329$, implying the non-equivalence between these circuits can be readily detected by the RH. 
\newpage
\begin{table*}[h!]
	\caption{Benchmark results for quantum circuit equivalence checking using VeriQBench dataset \cite{veriqbench}. }
	\label{tab_benchmarks} 
	\begin{tabular}{crrrrrr}
		\toprule
		Benchmark                &$n$ & \#gates & \#depth & $p_{\text{failure\_Y}}$ & $p_{\text{failure\_N}}$ & $t$ [$s$]\\  
		\hline
		2of5d2                   &7    &12   &8    & $1.511\times10^{-31}$  &0.322    &1.503    \\
		2of5d3                   &6    &137  &65   & $3.620\times10^{-31}$  &0.307    &0.466    \\
		3\_17tc                  &3    &6    &5    &$4.220\times10^{-32}$   &0.253    &0.068    \\
		4b15g\_4                 &4    &45   &24   &$1.113\times10^{-31}$   &0.279    &0.086     \\
		4b15g\_5                 &4    &15   &12   &$2.169\times10^{-32}$   &0.252    &0.073      \\
		4\_49\_fc                &4    &143  &72   &$3.267\times10^{-31}$   &0.279    &0.158     \\
		adder\_n4                &4    &4    &4    &$6.765\times10^{-32}$   &0.293    &0.100     \\
		adder\_n7                &7    &8    &5    &$1.531\times10^{-31}$   &0.328    &1.658     \\
		bv\_4                    &4    &12   &3    &$1.307\times10^{-31}$   &0.181    &0.107    \\
		bv\_5                    &5    &15   &4    &$1.843\times10^{-31}$   &0.191    &0.145    \\
		bv\_7                    &7    &21   &6    &$2.751\times10^{-31}$   &0.264    &2.369     \\
		bv\_9                    &9    &27   &8    &$3.777\times10^{-31}$   &0.250    &6.587      \\
		eff\_4\_4                &4    &19   &3    &$1.115\times10^{-31}$   &0.312    &0.184    \\
		eff\_5\_4                &4    &19   &3    &$1.025\times10^{-31}$   &0.243    &0.146   \\
		grover\_5                &5    &23   &4    &$2.269\times10^{-31}$   &0.291    &0.1815      \\
		grover\_7                &7    &32   &8    &$3.146\times10^{-31}$   &0.252    &2.8121     \\
		grover\_9                &9    &41   &12   &$3.999\times10^{-31}$   &0.167    &7.702    \\
		ham7tc                   &7    &143  &71   &$4.279\times10^{-31}$   &0.219    &7.520      \\
		hf\_6\_0\_9              &6    &155  &32   &$6.389\times10^{-31}$   &0.155    &0.805      \\
		hwb4\_11\_21             &4    &11   &10   &$7.354\times10^{-32}$   &0.312    &0.072      \\
		mod5d1                   &5    &8    &8    &$9.571\times10^{-32}$   &0.303    &0.116       \\
		nth\_prime4\_inc\_d1     &4    &72   &38   &$1.696\times10^{-31}$   &0.221    &0.126    \\
		nth\_prime7\_inc\_1427\_3172&7 &2513 &1566 &$3.358\times10^{-30}$   &0.329    &111.947      \\
		pe\_4                     &5   &18   &9    &$1.865\times10^{-31}$   &0.284    &0.181    \\
		pe\_5                     &6   &25   &12   &$2.465\times10^{-31}$   &0.172    &0.316    \\
		pe\_7                     &8   &42   &18   &$3.698\times10^{-31}$   &0.287    &3.711    \\
		pe\_8                     &9    &52   &21   &$4.324\times10^{-31}$  &0.160    &11.118 \\  
		qaoa\_6                   &6   &110  &9    &$4.668\times10^{-31}$   &0.260    &0.830      \\
		qcnn\_4                   &4   &66   &16   &$4.056\times10^{-31}$   &0.208    &0.264     \\
		qft\_4                    &4   &10   &5    &$1.178\times10^{-31}$   &0.243    &0.102    \\
		qft\_5                    &5   &15   &7    &$1.630\times10^{-31}$   &0.265    &0.188   \\
		qft\_7                    &7   &28   &11   &$2.602\times10^{-31}$   &0.260    &2.818     \\
		qft\_9                    &9   &45   &15   &$3.691\times10^{-31}$   &0.183    &11.671     \\
		rand\_cliff\_4\_AG        &4   &25   &9    &$1.490\times10^{-31}$   &0.166    &0.122   \\
		rand\_cliff\_7\_AG        &7   &83   &46   &$2.729\times10^{-31}$   &0.313    &5.268    \\
		rand\_cliff\_8\_AG        &8   &108  &59   &$3.505\times10^{-31}$   &0.270    &9.386   \\
		rand\_cliff\_9\_AG        &9   &107  &61   &$3.820\times10^{-31}$   &0.326    &22.744    \\
		rd32                      &4   &4    &4    &$7.213\times10^{-32}$   &0.252    &0.084   \\
		5QBT\_4CYC\_4GN\_1.0P2\_0 &4   &4    &4    &$7.154\times10^{-32}$   &0.315    &0.096      \\
		16QBT\_4CYC\_8GN\_1.0P2\_0&9   &8    &1    &$2.169\times10^{-31}$   &0.306    &5.236     \\
		20QBT\_4CYC\_8GN\_1.0P2\_0&5   &8    &3    &$1.150\times10^{-31}$   &0.250    &0.115    \\
		54QBT\_4CYC\_8GN\_1.0P2\_0&8   &8    &1    &$1.856\times10^{-31}$   &0.250    &1.983        \\
		quantum\_volume\_n5\_d5\_i0&5   &100  &15   &$8.247\times10^{-31}$   &0.298    &0.363      \\
		quantum\_volume\_n7\_d5\_i0&7   &150  &15   &$1.271\times10^{-30}$   &0.291   &9.102     \\
		quantum\_volume\_n9\_d5\_i0&9   &200  &15   &$1.716\times10^{-30}$   &0.155    &43.420     \\
		\bottomrule
	\end{tabular}
	\begin{tabular}{ccc}
		$n$: Number of qubits &  \#gates: Number of gates & \#depth: Circuit depth
	\end{tabular}
	\begin{tabular}{cc}
		$p_{\text{failure\_Y}}$: $p_{\text{failure}}$ for equivalent circuits &  $p_{\text{failure\_N}}$: $p_{\text{failure}}$ for non-equivalent circuits\\
	\end{tabular}
	\\
	\begin{tabular}{c}
		$t$: The total runtime for 100 parameter instances..
	\end{tabular}
	
\end{table*}

\end{document}